\documentclass[twocolumn,preprintnumbers,nofootinbib,prd,superscriptaddress,aps]{revtex4-1}

\usepackage[utf8]{inputenc} 
\usepackage{graphicx,amssymb,amsmath,amsthm,amsfonts,epstopdf,epsfig,epsf,times}
\usepackage[linktocpage]{hyperref}
\usepackage[usenames]{color}
\usepackage{epstopdf}
\usepackage{textcomp}
\usepackage{bm}
\usepackage{latexsym}
\usepackage{rotating}
\usepackage{hyperref}
\usepackage{color}
\usepackage{longtable}
\usepackage{enumerate}
\usepackage{tensor}
\usepackage{stmaryrd}
\usepackage[normalem]{ulem}
\usepackage{mathtools}
\usepackage{url}
\usepackage{multirow}
\usepackage{graphicx}
\usepackage[caption=false]{subfig}
\usepackage{mathtools}
\usepackage{verbatim}
\usepackage{soul,xcolor}
\setstcolor{red}

\definecolor{coolblack}{rgb}{0.0, 0.18, 0.39}
\definecolor{darkred}{rgb}{0.5,0,0}
\definecolor{darkgreen}{rgb}{0,0.5,0}
\definecolor{darkblue}{rgb}{0,0,0.5}
\definecolor{lapislazuli}{rgb}{0.15, 0.38, 0.61}
\definecolor{venetianred}{rgb}{0.78, 0.03, 0.08}
\definecolor{bleudefrance}{rgb}{0.19, 0.55, 0.91}
\definecolor{dogwoodrose}{rgb}{0.84, 0.09, 0.41}
\definecolor{dogwoodrose}{rgb}{0.84, 0.09, 0.41}
\definecolor{darkorgane}{rgb}{1,0.549,0}
\hypersetup{colorlinks=true, citecolor=darkblue, linkcolor=darkblue, 
urlcolor = darkblue}
\definecolor{olive}{rgb}{0.5, 0.5, 0.0}

\newcommand{\ben}{\begin{enumerate}}
\newcommand{\een}{\end{enumerate}}

\def\be{\begin{equation}}
\def\ee{\end{equation}}
\newcommand{\beq}{\begin{eqnarray}}
\newcommand{\eeq}{\end{eqnarray}} 
\newcommand{\ba}{\begin{align}}
\newcommand{\ea}{\end{align}}

\def\be{\begin{equation}}
\def\ee{\end{equation}}
	
\newcommand{\bea}{\begin{eqnarray}}
\newcommand{\eea}{\end{eqnarray}}

\usepackage{physics}

\begin{document}

\title{Love numbers and magnetic susceptibility of charged black holes}

\author{David Pereñiguez}
\affiliation{Instituto de Física Teórica UAM/CSIC, C/ Nicolás Cabrera, 13-15, C.U. Cantoblanco, E-28049 Madrid, Spain}
\affiliation{CENTRA, Departamento de F\'{\i}sica, Instituto Superior T\'ecnico -- IST, Universidade de Lisboa -- UL,
Avenida Rovisco Pais 1, 1049-001 Lisboa, Portugal}
\author{Vitor Cardoso}
\affiliation{CENTRA, Departamento de F\'{\i}sica, Instituto Superior T\'ecnico -- IST, Universidade de Lisboa -- UL,
Avenida Rovisco Pais 1, 1049-001 Lisboa, Portugal}
\affiliation{Niels Bohr International Academy, Niels Bohr Institute, Blegdamsvej 17, 2100 Copenhagen, Denmark}

\begin{abstract} 
The response of black holes to companions is of fundamental importance in the context of their dynamics and of gravitational-wave emission. 
Here, we explore the effect of charge on the static response of black holes. With a view to constraining broader setups, we consider 
charged geometries in an arbitrary number of spacetime dimensions $D\geq4$. 
Tensor tidal Love numbers are shown to follow a power law in the black hole temperature $\sim T_{H}^{2l+1}$, and thus vanish at extremality. In contrast, the black hole charge $Q$ excites new modes of polarisation in the vector sector that are otherwise not responsive in the neutral limit. In four dimensions, Love numbers and magnetic susceptibilities vanish for all values of the charge that respect the extremality bound. Using the theory of Fuchsian equations we are able to obtain analytical results in most cases, even beyond the hypergeometric instances. 
\end{abstract}

\maketitle
\section{Introduction}
The advent of gravitational-wave (GW) astronomy~\cite{LIGOScientific:2016aoc,Abbott:2020niy} and of very long baseline interferometry~\cite{EventHorizonTelescope:2019dse,GRAVITY:2020gka}
allows access to the hitherto invisible Universe~\cite{Barack:2018yly,Cardoso:2019rvt,Bertone:2018krk,Bar:2019pnz,Brito:2015oca}. In this vast landscape, compact objects such as black holes (BHs)
hold a tremendous discovery potential, allowing for unprecedented tests of General Relativity (GR) in the strong-field regime~\cite{Barack:2018yly,Berti:2016lat,Cardoso:2019rvt,Bertone:2018krk,Brito:2015oca,Seoane:2021kkk}: are BHs described by classical General Relativity~\cite{Chrusciel:2012jk} in vacuum, and up to which extent are matter effects important and measurable~\cite{Cardoso:2016ryw,Barausse:2014tra,Cardoso:2021wlq}? Do BHs exist and how can we {\it quantify} the presence of horizons in the spacetime~\cite{Cardoso:2017cqb,Cardoso:2019rvt}? 

The answer to the above questions requires an understanding of the dynamics of BH spacetimes in general setups, a notoriously difficult task. A key component in how BHs respond dynamically
lies in their deformability properties, encoded in so-called tidal Love numbers~(TLNs)~\cite{Murraybook,PoissonWill}. These leave a detectable imprint in the GW signal emitted by compact binaries in the late stages of their orbital evolution. An intriguing result in classical, vacuum GR concerns the {\it vanishing} of the TLNs of BHs~\cite{Binnington:2009bb,Damour:2009vw,Gurlebeck:2015xpa,LeTiec:2020spy,LeTiec:2020bos,Chia:2020yla}. The precise cancellation of the TLNs of BHs within Einstein's theory may pose a problem of ``naturalness''~\cite{Porto:2016pyg,Porto:2016zng,Rothstein:simons}, which can be argued to be as puzzling as the strong CP and the hierarchy problem in particle physics, or as the cosmological constant problem. The resolution of this issue in BH physics could lead to --~testable, since they would be encoded in GW data~--smoking-gun effects of new physics.

The above properties only hold in vacuum, while astrophysical BHs are surrounded by matter, even if dilute. 
Indeed, it was shown that such environmental effects can conspire to produce small but nonvanishing TLNs~\cite{Cardoso:2019upw}.
Other, light matter fields could arise in extensions of the Standard Model, or in higher dimensional theories~\cite{Dine:1982ah,Arvanitaki:2009fg,Marsh:2015xka}.
While their abundance could be negligible, it is unclear if their very existence contributes to nontrivial TLNs, but extra degrees of freedom, particularly scalar fields, can contribute with nonvanishing TLNs in some specific theories~\cite{Cardoso:2017cfl}.


Here, we address the following main question: what is the effect of charge and electromagnetic fields on the static polarisability of BHs? In particular, can charge excite new modes of static polarisation? Furthermore, we consider this in an arbitrary number of spacetime dimensions $D\geq 4$. This is a well-motivated setup for different reasons. First, the physics of higher-dimensional, charged BHs is a matter of interest \textit{per se}. In particular, these play a central role in the microscopic derivations of the Bekenstein--Hawking entropy~\cite{Strominger:1996sh,Maldacena:1996ky} as well as in the computation of its stringy corrections~\cite{Elgood:2020mdx,Elgood:2020nls}. Upon dimensional reduction, such BHs can also be relevant in astrophysics. While KK excitations do not seem reachable in astrophysical processes~\cite{Cardoso:2019vof}, in brane-world type reductions the extra-dimensions induce a definite signature in the BH frequency spectrum~\cite{Seahra:2004fg}. It is important to revisit these scenarios with focus on the static response. However, first one needs to understand the higher-dimensional degrees of freedom in more natural settings (e.g. $n$-dimensional spherical symmetry). Finally, from a more technical viewpoint, spacetime dimensionality $D$ can be seen as a regularisation parameter to obtain four-dimensional TLNs by taking $D\to4$, hence also understanding how special such parameter is in the space of possible values~\cite{Kol:2011vg,Hui:2020xxx}.

\section{Charged black holes in $D$ Dimensions}
We are interested in the static response of $D$-dimensional, asymptotically flat BHs which are charged under matter gauge fields. One of the simplest theories containing BHs fulfilling such requirements is Einstein--Maxwell theory in arbitrary spacetime dimension $D$. The field content is the metric $g_{AB}$ and a $U(1)$ gauge field $\mathcal{A}_{A}$, both subject to the action 
\begin{equation}
S[g,\mathcal{A}]=\frac{1}{2\kappa^{2}}\int d^{D}x\sqrt{g}\,R-\frac{1}{4}\int d^{D}x\sqrt{g} \,\mathcal{F}^{2}\,,\label{theory}
\end{equation}
where $\mathcal{F}=d\mathcal{A}$ is the field-strength and $\kappa^{2}$ the $D$-dimensional gravitational coupling. The equations of motion take the familiar form
\beq
G_{AB}&=&\kappa^{2}T_{AB}\,,\quad d\star \mathcal{F}=0\nonumber\\
T_{AB}&=&\mathcal{F}_{AC}\mathcal{F}_{B}^{\ C}-\frac{1}{4}g_{AB}\mathcal{F}^{2}\,.
\eeq
There is a large set of black objects solving these equations that are of interest in several contexts~\cite{Emparan:2008eg}. Here we are concerned with linear fluctuations on such spaces, which is a problem of significant complexity and hard to approach in various cases. An analysis of the perturbations based on harmonic decomposition is possible as long as the BH solutions enjoy enough structure~\cite{Ishibashi:2011ws,Kodama:2003kk}, and this is the only situation in which a complete description of the perturbations in arbitrary $D$ is known. Here we shall restrict to the BH solutions of~\eqref{theory} in which such analysis holds. 

Static, spherically-symmetric BHs of \eqref{theory} carrying electric charge are described by the Reissner--Nordstr\"{o}m--Tangherlini solutions~\cite{Reissner,Nordstrom,Tangherlini}. The metric and field strength read
\begin{equation}\label{solution}
ds^{2}=-fdt^{2}+\frac{dr^{2}}{f}+r^{2}d\Omega_{n}^{2},\ \ \ \ \mathcal{F}=E_{0}dt\wedge dr\,,
\end{equation}
where $f=f(r),\,E_0=E_0(r)$ and we find it convenient to define the dimension parameter
\be
n=D-2\,,
\ee
and
\begin{equation}
f=1-\frac{2M}{r^{n-1}}+\frac{Q^{2}}{r^{2n-2}},\ \ \ \ E_{0}=\frac{q}{r^{n}}, \ \ \ \ Q^{2}=\frac{\kappa^{2}q^{2}}{n(n-1)}\,,
\end{equation}
with $M$ and $Q$ the BH mass and charge (up to factors) respectively. The metric \eqref{solution} has Killing horizons relative to $k=\partial_{t}$ at
\begin{equation}
r_{\pm}^{n-1}=M\pm\sqrt{M^{2}-Q^{2}}\,.
\end{equation}
Consequently, the solution exhibits a regular event horizon at $r=r_{+}$ as long as the extremality bound $\vert Q\vert\leq M$ is preserved. In that case, the Hawking temperature of the BH is
\begin{equation}
T_{H}= \frac{n-1}{4\pi r_{+}^{n}}\left(r_{+}^{n-1}-r_{-}^{n-1}\right)\,.\label{temperature}
\end{equation}
When the extremality bound is saturated, the event and Cauchy horizons merge and $T_{H}=0$. On the other hand, as one approaches the neutral limit $Q=0$ the Cauchy horizon $r_{-}$ coalesces with the curvature singularity at $r=0$ and the solution reduces to Schwarzschild--Tangherlini \cite{Tangherlini}. We will see that this plays a crucial role for the master equations governing static perturbations. Whenever any of these two limits takes place, i.e. $Q=0$ or $T_{H}=0$, the equations become hypergeometric and Love numbers and magnetic susceptibilities are exactly solvable. For intermediate values of the BH charge the equations pick an extra pole (the Cauchy horizon) and are, therefore, less amenable. Nevertheless, we still manage to get exact results in most cases. In the following we derive the master equations governing static perturbations of \eqref{solution} for both the tensor and vector sectors.

\section{Perturbation theory}
A large class of BH spacetimes can be written as a warped product of an $n$-dimensional euclidean Einstein manifold $(\mathcal{K}^{n},\gamma_{ij})$ and an $m$-dimensional Lorentzian manifold $(\mathcal{N}^{m},g_{ab})$ ($i,j=1,..,n$ and $a,b=1,...,m$). The spacetime is $(n+m)$-dimensional with manifold structure $M=\mathcal{N}^{m}\times\mathcal{K}^{n}$ and, in adapted coordinates $x^{A}=(y^{a},z^{i})$, the metric takes the form
\begin{equation}\label{GeneralMetric}
ds^{2}=g_{ab}(y)dy^{a}dy^{b}+r^{2}(y)\gamma_{ij}(z)dz^{i}dz^{j}\,,
\end{equation}
where $r(y)$ is the warping factor defined as a function on $\mathcal{N}^{m}$. A metric with structure \eqref{GeneralMetric} is only compatible with energy-momentum tensors of 
the form
\begin{equation}
T_{ai}=0,\ \ \ T^{i}_{\ j}=P\delta^{i}_{\ j}\,,\label{P}
\end{equation}
where $P$ is a function on $\mathcal{N}^{m}$. Although such a spacetime is notably general, the fact that $\mathcal{K}^{n}$ is Einstein still allows an analysis of fluctuations based on harmonic decomposition. This is due to Kodama and Ishibashi (KI) who established a completely covariant and gauge-invariant approach to perturbation theory on these spaces \cite{Ishibashi:2011ws,Kodama:2003kk}.

In the KI formalism, taking advantage of the structure of Eq.~\eqref{GeneralMetric} one decomposes a general perturbation in tensor, vector and scalar sectors. After projection on the corresponding harmonics, Einstein's equations decouple in three sets of partial differential equations (PDEs) on $\mathcal{N}^{m}$, one for each sector. This holds for a general energy-momentum tensor, and the equations may be simplified by assuming its covariant conservation. Once a specific field content has been chosen, Einstein's equations are supplemented with the matter equations of motion. In the case of Einstein--Maxwell theory, the vector potential can already be decomposed in scalar $(\delta A_{a},a)$ and vector $A^{(1)}_{i}$ components\footnote{Equivalently, one may regard $\delta \mathcal{F}=d\delta \mathcal{A}$ as the basic variable and decompose it with respect to $\mathcal{K}^{n}$. This seems the most natural approach for matter fields of higher ranks.}
\begin{equation}
\delta \mathcal{A}=\delta A_{a} dy^{a}+\left(A^{(1)}_{i}+\hat{D}_{i}a\right)dz^{i}\,,\, \text{with} \ \ \hat{D}^{i}A^{(1)}_{i}=0\,,\label{vectordecomp}
\end{equation}
from which it follows that the matter tensor sector is empty in this theory. The final form of the equations is given in terms of a gauge-invariant basis of variables that can be constructed for each sector. The BHs of Einstein--Maxwell theory considered in this work, described by Eq.~\eqref{solution}, fall in the class of Eq.~\eqref{GeneralMetric} with $m=2$ and $\mathcal{K}^{n}=\mathbb{S}^{n}$. In the remaining of this work we adopt the KI formalism \cite{Ishibashi:2011ws,Kodama:2003kk} and focus on tensor and vector fluctuations on the background space \eqref{solution}. This is convenient because, on the one hand, it suffices to understand the behaviour of test fields as well as interacting gravitational and electromagnetic perturbations. In addition, the equations turn out to be simple enough so as to admit analytical results in several instances. A more thorough analysis including the scalar sector will be considered elsewhere.

\subsection{Master equations and their static limit: tensor sector}
A general tensor perturbation is generated by just two gauge-invariant variables $(H_{T},\tau_{T})$ \cite{Ishibashi:2011ws,Kodama:2003kk},
\begin{equation}
h_{ij}=2r^{2}H_{T}\mathbb{T}_{ij}, \ \ \ \delta T_{ij}=r^{2}\left(\tau_{T}+2 P H_{T}\right)\mathbb{T}_{ij}\,,
\end{equation}
where $\mathbb{T}_{ij}$ are the tensor harmonics on $\mathbb{S}^{n}$ satisfying
\begin{equation}
\left(\hat{D}^{k}\hat{D}_{k}+k_{t}^{2}\right)\mathbb{T}_{ij}=0, \ \ \ \mathbb{T}^{i}_{\ i}=0=\hat{D}^{j}\mathbb{T}_{ji}\,,
\end{equation}
with spectrum 
\be
k_{t}^{2}=L(L+n-1)-2\,,\qquad L=2,...\label{eq:harmonic_tensor}
\ee
Furthermore, the Maxwell field-strength $\delta F$ does not contribute to the tensor part of the energy-momentum tensor, 
\begin{equation}
\tau_{T}=0\,,
\end{equation}
and the Einstein--Maxwell equations reduce to a single PDE on $\mathcal{N}^{2}$ for $H_{T}$,
\begin{equation}
\square H_{T}+\frac{n}{r}Dr\cdot DH_{T}-\frac{k_{t}^{2}+2}{r^{2}}H_{T}=0\,.\label{eqscalar}
\end{equation}
As noted by KI~\cite{Ishibashi:2011ws,Kodama:2003kk}, Eq.~\eqref{eqscalar} turns out to be the same as that satisfied by a test, massless scalar field on our background if $k_{t}^{2}$ is appropriately identified with the angular momentum number. Therefore, the tensor sector can also be used to infer properties of test fields on \eqref{solution}. 

After a field redefinition $H_{T}=r^{-n/2}\phi$ to get rid of the term $\sim Dr\cdot DH_{T}$ the master equation becomes
\begin{equation}
\left(\square+V\right)\phi=0\,,\label{mastdyn}
\end{equation}
with
\begin{equation}
V=\frac{n(3n-2)}{4}\frac{Q^{2}}{r^{2n}}-\frac{4k_{t}^2+8+n^{2}-2n}{4r^{2}}-\frac{n^{2}}{2}\frac{M}{r^{n+1}}\,.
\end{equation}
We are interested in the static solutions of this equation, that is, solutions satisfying $\pounds_{k}\phi=0$ where $k$ is the static time-like Killing vector of \eqref{solution}. Either in Schwarzschild or Eddington--Finkelstein coordinates, this translates into the requirement that $\phi$ is a function of $r$ only, $\phi=\phi(r)$. When specialised for a static perturbation,
Eq.~\eqref{mastdyn} becomes an ODE of Fuchsian type with four regular singular points: infinity, the event horizon, the Cauchy horizon and the singularity. Therefore, it can be cast in Heun's form \cite{Kristensson,NIST:DLMF}. To see this, we first introduce the dimensionless variable
\begin{equation}\label{z}
z=\left(\frac{r_{+}}{r}\right)^{n-1}\,.
\end{equation}
Then, after a field redefinition 
\begin{equation}
H_{T}(z)=r(z)^{-n/2}z^{\frac{2l(n-1)+n-2}{2(n-1)}}\Psi(z)\,,
\end{equation}
the master equation becomes of Heun's type, 
\beq
&&\Psi ''+ \left(\frac{\gamma}{z}+\frac{\delta}{z-1}+\frac{\eta}{z-z_{c}}\right)\Psi '\nonumber\\
&+&\frac{\alpha \beta \left(z-h\right)}{z(z-1) \left(z-z_{c}\right)}\Psi=0\,,\label{mastertensor}
\eeq
where primes stand for derivatives with respect to $z$ and with coefficients
\beq
z_{c}&=&\cot ^2\left(\frac{\epsilon }{2}\right),\ \ \gamma=2(l+1),\ \ \delta=1, \ \ \eta=1\,,\nonumber\\
\alpha&=&2+l,\ \ \beta=1+l,\ \  h=\frac{(l+1) \csc ^2\left(\frac{\epsilon }{2}\right)}{l+2}\,.\label{mastertensorcoefs}
\eeq
This equation depends on two dimensionless parameters, $\epsilon$ and $l$, defined as
\begin{equation}
l=\frac{L}{n-1},\ \ \ \ \sin\epsilon=\frac{Q}{M}\,,\label{leps}
\end{equation}
where $L$ is the harmonic number defined by \eqref{eq:harmonic_tensor}. The extremality bound dictates $\vert Q\vert \leq M$, and without loss of generality we can restrict to $\epsilon\in[0,\pi/2]$ with neturality and extremality lying at $0$ and $\pi/2$, respectively. Equation \eqref{mastertensor} has regular poles at $z=0,1,z_{c},\infty$, corresponding respectively to infinity, the event and Cauchy horizons and the singularity.

\begin{figure}[t!]
\includegraphics[width=0.50\textwidth]{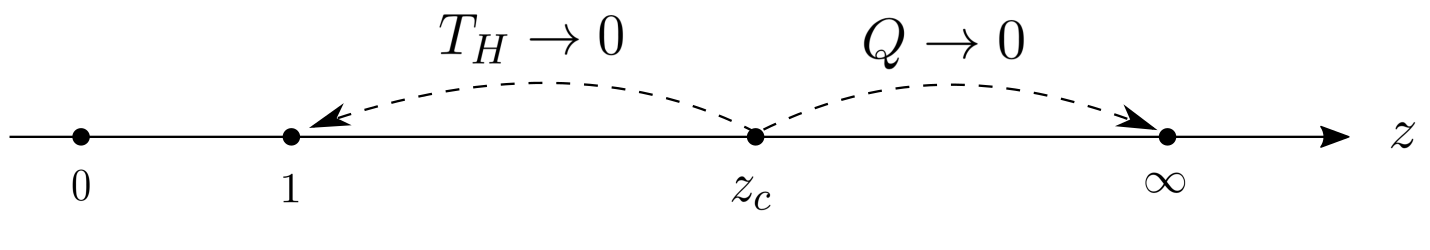}
\caption{\small{Singularity structure of the master equations. The regular singular point at the Cauchy horizon $z=z_{c}$ coalesces with those at the event horizon $z=1$ and spacetime singularity $z=\infty$ in the extremal and neutral limits, respectively.}}
\label{fig:poles}
\end{figure}

It is interesting to specialize the general equation \eqref{mastertensor} to neutral and extremal cases. The regular singularity at the Cauchy horizon $z_{c}$ collides with that on the event horizon as $T_{H}\to0$, while in the neutral limit $Q\to0$ it merges with the spacetime curvature singularity (see Figure \ref{fig:poles}). Quite interestingly, in none of these limits the merging produces an irregular singularity. Instead, one has three regular singularities at infinity $z=0$, the horizon $z=1$ and the curvature singularity $z=\infty$. Consequently, the equation becomes of hypergeometric type and in such cases one can use the theory of hypergeometric functions to obtain analytically the response parameters, as discussed in \cite{Kol:2011vg,Hui:2020xxx} for the neutral limit. We will find the same pole structure in the vector sector.

Explicitly, in the neutral case equation \eqref{mastertensor} can be immediately evaluated at $\epsilon=0$ giving the hypergeometric equation
\begin{equation}
z(1-z)\Psi''+\left[c-(a+b+1)z\right]\Psi'-a b \Psi=0\,,\label{mastertensorlimits}
\end{equation}
with coefficients 
\begin{equation}
a=b=l+1, \ \ \ c=2(l+1)\,.\label{mastertensorcoefsneut}
\end{equation}
This coincides with the equation obtained in Ref.~\cite{Hui:2020xxx} for the tensor degree of freedom. In the extremal case $\epsilon=\pi/2$, after a field redefinition
\begin{equation}
\Psi(z)=(1-z)^{l}\psi(z)\,,\label{tensvarext}
\end{equation}
one obtains again an hypergeometric equation \eqref{mastertensorlimits} for $\psi(z)$, now with parameters
\begin{equation}
a=c=2(l+1)\,,\, b=2l+1\,.\label{mastertensorcoefsext}
\end{equation}
In sum, we have found that a static tensor perturbation is governed by Heun's equation \eqref{mastertensor} with coefficients given in \eqref{mastertensorcoefs}. In the neutral and extremal limits, it reduces to an hypergeometric equation \eqref{mastertensorlimits} with coefficients given in \eqref{mastertensorcoefsneut} and \eqref{mastertensorcoefsext}, respectively. In the following section we will discuss solutions to these equations and obtain the associated response parameters.  

\subsection{Master equations and their static limit: vector sector}
The vector sector of a general perturbation is composed of~\cite{Ishibashi:2011ws,Kodama:2003kk}
\begin{align}\label{vectharmdec1}
h_{ai}&=h^{(1)}_{a}\mathbb{V}_{i}\,,\quad h_{ij}=-2k_{v}h^{(1)}_{T}\mathbb{V}_{ij}\,,\\
\delta T_{ai}&=T^{(1)}_{a}\mathbb{V}_{i}\,,\quad \delta T_{ij}=-2k_{v}T^{(1)}_{T}\mathbb{V}_{ij}\,,\label{vectharmdec2}
\end{align}
where the vector harmonics $\mathbb{V}_{i}$ satisfy
\beq
&&\left(\hat{D}^{j}\hat{D}_{j}+k_{v}^{2}\right)\mathbb{V}_{i}=0\,,\, \hat{D}^{i}\mathbb{V}_{i}=0\,,\, \mathbb{V}_{ij}:=-\frac{1}{k_{v}}\hat{D}_{(i}\mathbb{V}_{j)}\,,\nonumber \\
&&k_{v}^{2}=L(L+n-1)-1\,,\quad L=1,2,...
\eeq
Excluding the special harmonic case $L=1$, a basis of gauge-invariant variables in $\mathcal{N}^{2}$ is
\beq
F^{(1)}_{a}&=&\frac{1}{r}\left(h^{(1)}_{a}-r^{2}D_{a}\left(\frac{h^{(1)}_{T}}{r^2}\right)\right)\,,\label{gaugeinv}\\
\tau^{(1)}_{a}&=&\frac{1}{r}\left(T^{(1)}_{a}-P h^{(1)}_{a}\right)\,,\\
\tau_{T}&=&\frac{2k_{v}}{r^{2}}\left(-T^{(1)}_{T}+Ph^{(1)}_{T}\right)\,.
\eeq
There are two Einstein equations for this sector plus one coming from conservation of $T_{AB}$, $\delta(\nabla^{M}T_{MA})=0$. The latter can be combined with one of the Einstein equations to give an integrability condition, which allows one to trade $F_{a}$ by a function $\Omega$ satisfying
\beq
D_{a}\Omega&=&\epsilon_{ac}\left(r^{n-1}F^{c}-2\frac{\kappa^{2}}{m_{V}}r^{n+1}\tau^{c}\right)\,,\label{integrabilityvector}\\
m_{V}&=&k_{v}^{2}-(n-1)=(L-1)(L+n)\,.
\eeq
Notice that $m_{V}=0$ only for the special harmonic $L=1$ that we consider separately. In addition, the vector sector of Maxwell's field is generated by a single gauge-invariant function $A$ on $\mathcal{N}^{2}$,
\begin{equation}
\delta \mathcal{A}=A\mathbb{V}_{i}dz^{i}, \ \ \ \tau_{a}=-\frac{q}{r^{n+1}}\epsilon_{ab}D^{b}A,\ \ \ \tau_{T}=0\,.
\end{equation}
In terms of the gauge-invariant functions $(\Omega, A)$ on $\mathcal{N}^{2}$, the Einstein and Maxwell equations are reduced to a pair of coupled PDEs
\begin{align}\label{v1}
&r^{n}D^{a}\left(\frac{D_{a}\Omega}{r^{n}}\right)-\frac{m_{V}}{r^{2}}\Omega=-\frac{2\kappa^{2}}{m_{V}}r^{n}\epsilon^{ab}D_{a}\left(r \tau_{b}\right)\,,\\
&\frac{1}{r^{n-2}}D_{a}\left(r^{n-2}D^{a}A\right)-\frac{k_{v}^{2}+n-1}{r^{2}}A=\frac{q m_{V}}{r^{2n}}\Omega \label{v2}\,.
\end{align}
Introducing the field redefinitions
\begin{equation}
\phi_{\pm}=a_{\pm}r^{-n/2}\left(\Omega-\frac{2\kappa^{2} q}{m_{V}}A\right)+b_{\pm}r^{\frac{n-2}{2}}A\,,\label{dof}
\end{equation}
we find that equations \eqref{v1} and \eqref{v2} decouple if
\begin{align}
(a_{+},b_{+})&=\left(\frac{Q m_{V}}{\Delta +M \left(n^2-1\right)}\sigma_{(+)},\frac{Q}{q}\sigma_{(+)}\right)\,,\label{plus}\\ 
(a_{-},b_{-})&=\left(\sigma_{(-)},- \frac{2  \kappa ^2 q}{\Delta +M \left(n^2-1\right)}\sigma_{(-)}\right)\,,\label{minus}
\end{align}
where $\sigma_{(\pm)}$ are any two (non-zero) constants and the positive constant $\Delta$ satisfies
\begin{equation}
\Delta^{2}=M^2 \left(n^2-1\right)^2+2 n(n-1)m_{V} Q^{2} \,.
\end{equation}
With this, $\phi_{\pm}$ satisfy master equations of the form
\begin{equation}
\left(\square+V_{\pm}\right)\phi_{\pm}=0\,,\label{mastervects}
\end{equation}
with
\beq
V_{\pm}&=&-\frac{k_{v}^{2}+1+n^{2}/4-n/2}{r^{2}}-\frac{n(5n-2)Q^{2}/4}{r^{2n}}\nonumber\\
&-&\frac{-(n^{2}+2)M/2\pm\Delta}{r^{n+1}}\,.\label{potentials}
\eeq

A comment here is in order. This derivation of Eq.~\eqref{mastervects} reproduces that in Ref.~\cite{Ishibashi:2011ws,Kodama:2003kk} with the difference that the decoupling parameters \eqref{plus} and \eqref{minus} are defined only up to their global factors. This is due to the fact that the general solution must depend on two independent amplitudes. In the neutral background these are clearly associated to the gravitational and electromagnetic fluctuations. However, when the BH is charged, such fluctuations couple and the independent amplitudes refer to the modes $\phi_{\pm}$ that contain fixed proportions of gravitational and electromagnetic contributions. This fact will be important for the definition of vector Love numbers and magnetic susceptibilities. Lastly, notice that on the neutral background $\phi_{-}$ and $\phi_{+}$ reduce to the standard gravitational and electromagnetic master variables respectively, so it may still be sensible to regard them as the gravitational and electromagnetic degrees of freedom even in the charged case. 

We are interested in static solutions of \eqref{mastervects}. In terms of the new variables $\Psi_{\pm}$,
\begin{equation}
\phi_{\pm}=z^{\frac{2 l (n-1)+n-2}{2 (n-1)}}\Psi_{\pm}\,,\label{Psivec}
\end{equation}
we obtain once again Heun's differential equation~\eqref{mastertensor},
but now with parameters given by
\beq
z_{c}&=&\cot ^2\left(\frac{\epsilon }{2}\right)\,,\,\, \alpha=l+3+\frac{1}{n-1}\,,\, \beta=l-\frac{1}{n-1}\,,\,\nonumber \\
\gamma&=&2(l+1)\,,\, \delta=\eta=1\,,\nonumber\\
h_{\pm}&=&\frac{\csc ^2\left(\frac{\epsilon }{2}\right) \left(1+2 l (n-1)^2(l+2)+n^2-4 n\pm\tilde{\Delta} \right)}{2 \left(2+l^2 (n-1)^2+3 l (n-1)^2-3 n\right)}\,.\label{mastervectorparams}
\eeq
The dimensionless variable $z$ and parameters $l$ and $\epsilon$ are given by \eqref{z} and \eqref{leps}, respectively, while $\tilde{\Delta}:=\Delta/M$. 

This equation has the same pole structure as the master equation of the tensor sector, with regular singularities at infinity $z=0$, event horizon $z=1$, Cauchy horizon $z=z_{c}$ and curvature singularity $z=\infty$. Just as in the tensor case, the Cauchy horizon $z_{c}$ merges with the singular points at the event horizon and the curvature singularity in the extremal and neutral limits, respectively, leading in both cases to an hypergeometric equation (see Figure \ref{fig:poles}). The equations for the neutral case are obtained just by evaluating \eqref{mastertensor} at $\epsilon=0$, and have hypergeometric form
\begin{equation}
z(1-z)\Psi''_{\pm}+\left[c-(a_{\pm}+b_{\pm}+1)z\right]\Psi'_{\pm}-a_{\pm} b_{\pm} \Psi_{\pm}=0\,,\label{neutvect}
\end{equation}
with parameters $c=2(l+1)$ and
\beq
a_{+}&=&a_{-}+1=l-\frac{1}{n-1}+1\,,\label{extneut1}\\
b_{+}&=&b_{-}-1=l+\frac{1}{n-1}+1\,.\label{extneut}
\eeq
Again, we find agreement with previous results for the neutral case~\cite{Hui:2020xxx}. The extremal limit is a bit more involved. After a field redefinition
\begin{equation}\label{extmastvar}
\Psi_{\pm}=(1-z)^{\frac{1}{2}\left(\frac{\Sigma_{\pm}}{n-1}-1\right)}\psi_{\pm}\,,
\end{equation}
where 
\begin{equation}
\Sigma_{\pm}=\sqrt{(n-1) (5 n+3)+4 \left(m_{V}\pm\tilde{\Delta}\right)}\,,
\end{equation}
and introducing the symbol
\begin{equation}\label{symbol}
S_{(\rho,\sigma)}=\frac{\Sigma_{\rho}+\sigma(3 n-1)}{2 (n-1)}\ \ \ \text{with} \ \ \ \rho,\sigma=\pm\,,
\end{equation}
the equations for $\psi_{\pm}$ take hypergeometric form with $c=2(l+1)$ and
\begin{equation}\label{extparam}
a_{\pm}=1+l+S_{(\pm,+)},\ \ \ b_{\pm}=1+l+S_{(\pm,-)}\,.
\end{equation}

Lastly, we consider the special harmonic mode. This corresponds to the case that $\mathbb{V}_{i}$ is a Killing vector field in $\mathcal{K}^{n}$, i.e. $\mathbb{V}_{ij}=0$. For $\mathcal{K}^{n}=\mathbb{S}^{n}$ this happens only if $m_{V}=(L-1)(L+n)=0$, i.e. $L=1$ \cite{Ishibashi:2011ws,Kodama:2003kk}. The projection of the perturbation into this harmonic, unlike the general one \eqref{vectharmdec1} and \eqref{vectharmdec2}, is composed of just 
\begin{align}
h_{ai}&=h^{(1)}_{a}\mathbb{V}_{i}\,,\, \delta T_{ai}=T^{(1)}_{a}\mathbb{V}_{i}\,.
\end{align}
The gauge-invariant variables in this case are
\begin{align}
F_{ab}&=rD_{a}\left(\frac{h^{(1)}_{b}}{r^{2}}\right)-rD_{b}\left(\frac{h^{(1)}_{a}}{r^{2}}\right)\,,\\
\tau_{a}&=\frac{1}{r}\left(T^{(1)}_{a}-P h^{(1)}_{a}\right)=\frac{-q}{r^{n+1}}\epsilon_{ab}D^{b}A\,,
\end{align}
and $F_{ab}$ can be solved exactly as \cite{Ishibashi:2011ws,Kodama:2003kk}
\begin{equation}
F=q\frac{2\kappa^2}{r^{n+1}}A-\frac{2\kappa^{2}}{r^{n+1}}\tau_{0}\,,
\end{equation}
where $F=(1/2)\epsilon^{ab}F_{ab}$ and $\tau_{0}$ is an arbitrary integration constant. It follows that the gravitational special mode is non-dynamical. In particular, $\tau_{0}$ generates a small rotation so restricting to a static background requires setting $\tau_{0}=0$. In terms of $\phi_{+}=r^{\frac{n-2}{2}}A$, Maxwell's equation reduces precisely to the ``$+$'' equation in \eqref{mastervects} with $L=1$.

In sum, we have found that static vector perturbations are governed by equations of Heun's type \eqref{mastertensor} with parameters \eqref{mastervectorparams}. In the neutral and extremal limits these become hypergeometric, with parameters \eqref{extneut1}-\eqref{extneut} and \eqref{extparam}, respectively. The special harmonic is recovered by just setting $L=1$ in the electromagnetic mode ($+$) and disregarding the gravitational one ($-$). In the following section we discuss static solutions to these equations and obtain the associated response parameters. 

\section{Static response}
The original works that established the vanishing of BH Love numbers in four dimensions, both in neutral~\cite{Binnington:2009bb} and charged~\cite{Cardoso:2017cfl} cases, followed an approach based on a full GR computation. Recently, the authors in Ref.~\cite{Hui:2020xxx} considered also this point of view to compute the static response of fields with integer spin, $0,1$ and $2$, fluctuating on a neutral Schwarzschild--Tangherlini background. Along the lines of \cite{Kol:2011vg}, they also showed that response parameters obtained in that way can be regarded as coefficients in a worldline effective action associated to the BH, thus clarifying some concerns about ambiguities in the definition of Love numbers~\cite{Fang:2005qq,Gralla:2017djj}. All these motivates us to adopt a full GR approach to study the static response of charged BHs in arbitrary $D$.
\subsection{Tensor Love numbers}\label{DefLove}
The parameters governing the static response of a system to a tidal field can be obtained by inspection of the solutions at infinity. Consider first a tensor perturbation on \eqref{solution}, which is described by \eqref{mastertensor}. From the standard theory Fuchsian equations, in a neighbourhood of $z=0$ the general solution has the form \cite{Kristensson}
\begin{equation}\label{gensol}
\Psi(z)=A \Psi_{\rm resp}(z)+B\left(z^{-2l-1}\Psi_{\rm tidal}(z)+R \Psi_{\rm resp}(z) \ln z\right)\,.
\end{equation}
Here, $A$ and $B$ are arbitrary constants multiplying two linearly independent solutions. The first one, $\Psi_{\rm resp}(z)$, is analytic at $z=0$ and without loss of generality we choose to normalise it as $\Psi_{\rm resp}(z)=1+O(z)$. The second solution contains, in general, a logarithmic term where $R$ is some constant and $\Psi_{\rm tidal}(z)$ is another analytic function at $z=0$ that we chose to normalise as $\Psi_{\rm tidal}(z)=1+O(z)$. Of course, the indices of our equation at $z=0$ are $0$ and $-(2l+1)$, and the latter quantity serves as a discriminant between qualitatively different cases:

\noindent $\bullet$  $2l+1\notin \mathbb N$: In this case the Frobenius solutions associated to each index at $z=0$ are linearly independent and one has $R=0$. After imposing regularity at the horizon $z=1$ the relative normalisation between $A$ and $B$ gets fixed,
\begin{equation}
\Psi(z)=B\left(k \Psi_{\rm resp}(z)+z^{-2l-1}\Psi_{\rm tidal}(z)\right)\,.\label{enn}
\end{equation}
The growing mode at infinity $\sim z^{-2l-1}$ has the interpretation of an external tidal field while $\Psi_{\rm resp}(z)$, which is regular at $z=0$, is the response of the system. The parameter $k$ is the (dimensionless) tidal Love number, which is precisely the quantity controlling the fall-off induced by the tidal field. Since it is completely determined by the requirement of regularity at the horizon and does not depend on the amplitude of the tidal field, the Love number $k$ is an intrinsic property of the BH.

\noindent $\bullet$ $2l+1\in \mathbb N$: In general, the second solution exhibits a logarithmic term, so the constant $R$ may not vanish. Again, regularity at the horizon $z=1$ fixes the relative normalisation between $A$ and $B$,
\begin{equation}
\Psi(z)=B\left(k \Psi_{\rm resp}(z)+z^{-2l-1}\Psi_{\rm tidal}(z)+R \Psi_{\rm resp}(z) \ln z\right)\,.\label{solR}
\end{equation} 
However, unlike the case where $2l+1\notin \mathbb N$, now the quantity $k$ is ambiguous due to power mixing. From the regular solution \eqref{solR}, there is no natural way of telling apart which contribution to the power series comes from the response and which from the tidal field. In particular, $k \Psi_{\rm resp}(z)$ can be completely absorbed order by order in the term $z^{-2l-1}\Psi_{\rm tidal}(z)$. Similar observations where noted in \cite{Hui:2020xxx}. The invariant piece of information here is $R$. Furthermore, as shown in \cite{Kol:2011vg} and discussed in \cite{Hui:2020xxx} the logarithmic term corresponds to a classical RG running of the induced response which is characterised by $R$, so we shall take $R\ln z$ as the response ``parameter'' in this case. Nevertheless, there is a remarkable exception within the case $2l+1\in \mathbb N$. It may be that \eqref{mastertensor} admits a second solution where $R=0$ and $\Psi_{\rm tidal}(z)$ is a polynomial of degree $\leq2l+1$. This purely growing mode is a tidal field and, furthermore, being just a terminating series in $z$ it is precisely the solution that is regular on the horizon $z=1$. It follows that the Love number is zero in this case\footnote{There is some discussion on whether this argument can be applied to the rotating case~\cite{LeTiec:2020spy,Chia:2020yla}.}. As shown below, this is exactly what happens in $D=4$.

Notice that this definition of Love numbers is in complete analogy with those in the literature in several contexts \cite{Binnington:2009bb,Cardoso:2017cfl,Kol:2011vg,Emparan:2017qxd,Hui:2020xxx} and, in particular, it reduces exactly to that of \cite{Hui:2020xxx} for the neutral BH. In the following we compute the tensor Love numbers for neutral and extremal limits separately, and then consider the case of finite charge and temperature.

\subsubsection{Neutral and extremal limits}
For vanishing BH charge $Q=0$ static tensor perturbations are governed by the hypergeometric equation \eqref{mastertensorlimits} with parameters \eqref{mastertensorcoefsneut}. Writing the general solution in terms of hypergeometric functions and using the connection formulas between Kummer's solutions, the authors in \cite{Hui:2020xxx} computed the response parameters defined as in the previous section. We list them here for completeness, 
\begin{equation}\label{neutraltensorlove}
k^{(\rm neut)}_{\rm tensor}=\begin{cases} \frac{2l+1}{2\pi}\frac{\Gamma\left(l+1\right)^{4}}{\Gamma\left(2l+2\right)^{2}}\tan{(\pi l)}& l\notin\mathbb{N},\frac{1}{2}\mathbb{N}\\  \frac{(-1)^{2l}\Gamma\left(l+1\right)^{2}}{(2l)!\left(2l+1\right)!\Gamma\left(-l\right)^{2}}\ln z& l\in\frac{1}{2}\mathbb{N}\\ 0 & l\in\mathbb{N}\end{cases}
\end{equation}
and notice that the only relevant case in $D=4$ is $l\in\mathbb{N}$. In the extremal case the static tensor perturbation $\psi$ in \eqref{tensvarext} is likewise subject to an hypergeometric equation, but now with parameters \eqref{mastertensorcoefsext}. Such equation turns out to admit a remarkably simple general solution for all $l$,
\begin{equation}
\psi(z)=\frac{A}{(1-z)^{2l+1}}+\frac{B}{z^{2l+1}}\,,\label{sol}
\end{equation}
with $A$ and $B$ arbitrary constants. Clearly, imposing regularity at the horizon $z=1$ fixes $A=0$, thus leaving just a pure tidal field $\psi(z)\sim z^{-2l-1}$. This leads to the interesting result that tensor Love numbers vanish at extremality in any number of spacetime dimensions,
\begin{equation}
k^{(\rm ext)}_{\rm tensor}=0\,.\label{EL}
\end{equation}

\subsubsection{Finite charge and temperature}
\begin{figure}[t!]
		\includegraphics[width=0.45\textwidth]{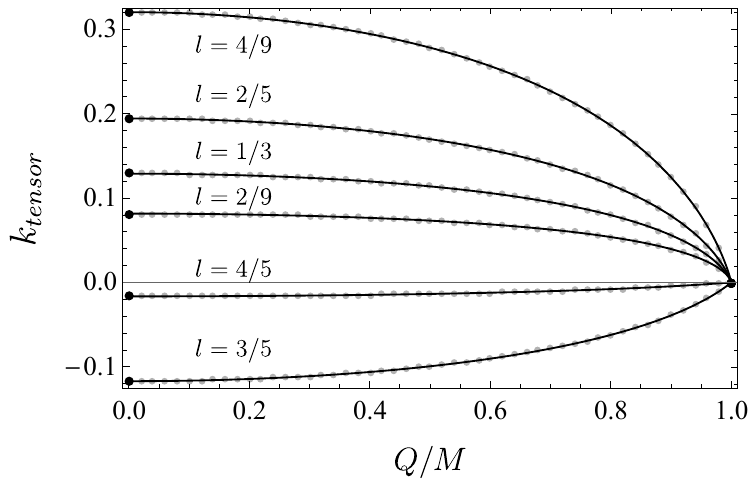}
		\caption{Tensor TLNs for some values of $l=L/(n-1)$ in the generic case $2l+1\notin\mathbb{N}$. For $n=6,10$, we represent $L=2,3,4$. Gray dots are the numerical values, solid black dots at the edges are the analytic predictions at neutrality $Q=0$ and extremality $Q=M$, and the solid black lines correspond to the analytic formula \eqref{LoveKGen}. We observe that \eqref{LoveKGen} is indeed in perfect agreement with both analytic and numeric results.}
		\label{fig:tensor}
\end{figure}

For intermediate charges $0<Q<M$, the Cauchy horizon introduces an additional pole in the master equation, which becomes of Heun's type \eqref{mastertensor}. Unfortunately, the latter is not as symmetric as the hypergeometric equation, so no analogue of Kummer's solutions exist and connection formulas are not available in general \cite{Bateman:100233,Kristensson,NIST:DLMF}. Thus, it is not clear how to write suitably the analytic prolongation of a solution, say, from a neighbourhood of $z=0$ to a neighbourhood of $z=1$\footnote{See \cite{Bonelli:2021uvf,Bonelli:2022ten} for recent progress in tackling this issue for Heun's confluent equation, which is the relevant ODE for oscillating perturbations in Kerr's black hole.}. This makes it difficult to obtain the response parameters proceeding as in the neutral and extremal limits. Rather remarkably, though, for tensor perturbations it is possible to obtain analytical results for all $l$. Consider first the degenerate case $l\in\frac{1}{2}\mathbb{N}$. After choosing the normalisation of $\Psi_{tidal}(z)$ as $\Psi_{tidal}(z)=1+O(z)$, equation \eqref{mastertensor} applied to the second solution of \eqref{gensol} fixes $R$ completely and it is possible to obtain its exact value after solving just a few orders. Furthermore, the result can be written in closed form 
\beq
R_{\rm tensor}&=&R^{(\rm neut)}_{\rm tensor}\left(\frac{\cos\epsilon}{\cos^{2}\left(\epsilon/2\right)}\right)^{2l+1}\nonumber\\
&=&R^{(\rm neut)}_{\rm tensor}\left(\frac{4\pi r_{+}}{n-1}T_{H}\right)^{2l+1} \ \ \ \ \left(l\in\frac{1}{2}\label{LoveRtens}\mathbb{N}\right)\,,
\eeq
where $r_{+}$ and $T_{H}$ are the radius and temperature of the BH, and $R^{\rm (neut)}_{\rm tensor}$ is the coefficient in front of the logarithm in the neutral case \eqref{neutraltensorlove}. Notice that \eqref{LoveRtens} vanish at extremality, $T_{H}=0$, as expected from the result in \eqref{EL}. For $l\in\mathbb{N}$ we find that the second solution is just $z^{-2l-1}\Psi_{\rm tidal}(z)$, with no logarithmic term, where $\Psi_{\rm tidal}(z)$ is a polynomial of degree $l$, so
\begin{equation}
k_{\rm tensor}=0 \ \ \ \left(l\in\mathbb{N}\right)\,.\label{LovekDeg}
\end{equation} 
This is the only relevant case for $D=4$, where $l$ takes values just in $\mathbb{N}$. Tensor perturbations do not exist in four dimensions, but due to the close relation between the tensor sector and (massless) scalar fields, the result \eqref{LovekDeg} shows that 4$D$, electrically charged BHs do not polarise under tidal fields of scalar type. Finally, for $l\notin\mathbb{N},\frac{1}{2}\mathbb{N}$ with no connection formulas available it is most likely that the only way of obtaining the Love numbers at finite $Q$ and $T_{H}$ is numerically. However, in views of the results \eqref{LoveRtens} and \eqref{LovekDeg} it is very tempting to try with
\begin{equation}
k_{\rm tensor}=k^{\rm (neut)}_{\rm tensor}\left(\frac{4\pi r_{+}}{n-1}T_{H}\right)^{2l+1}  \ \ \ \ \left(l\notin \mathbb{N},\frac{1}{2}\mathbb{N}\right)\,.\label{LoveKGen}
\end{equation}
where $k^{\rm (neut)}_{\rm tensor}$ is the neutral Love number shown in \eqref{neutraltensorlove}. We compared this expression with the numerical results obtained for $k_{\rm tensor}$ and have found exact agreement. In Figure \ref{fig:tensor} we illustrate this for various values of $l$. This confirms the validity of \eqref{LoveKGen}, although a rigorous proof is still desirable.

We conclude that the tensor Love numbers of a charged BH of radius $r_{+}$ at temperature $T_{H}$ are 
\beq
k_{\rm tensor}&=&k^{\rm (neut)}_{\rm tensor}\left(\frac{4\pi r_{+}}{n-1}T_{H}\right)^{2l+1}\nonumber\\
&=&\begin{cases} \frac{2l+1}{2\pi}\frac{\Gamma\left(l+1\right)^{4}}{\Gamma\left(2l+2\right)^{2}}\tan{(\pi l)}\left(\frac{4\pi r_{+}}{n-1}T_{H}\right)^{2l+1}& l\notin\mathbb{N},\frac{1}{2}\mathbb{N}\\  \frac{(-1)^{2l}\Gamma\left(l+1\right)^{2}}{(2l)!\left(2l+1\right)!\Gamma\left(-l\right)^{2}} \left(\frac{4\pi r_{+}}{n-1}T_{H}\right)^{2l+1} \ln z& l\in\frac{1}{2}\mathbb{N}\\ 0 & l\in\mathbb{N}\end{cases}\label{TensorLove}
\eeq
It is clear that these vanish at extremality, $T_{H}=0$, thus recovering \eqref{EL}, and reduce to those obtained in \cite{Hui:2020xxx} for $Q=0$ (see \eqref{neutraltensorlove}). At this point it is natural to wonder how general the vanishing of Love numbers at extremality is. In the following section we show that vector Love numbers and magnetic susceptibilities do not vanish at $T_{H}=0$. Instead, BHs become significantly more polarised as one approaches the extremality bound. 

\subsection{Vector Love numbers and magnetic susceptibility}
Response parameters $k_{\pm}$ can be defined for the master variables of the vector sector $\Psi_{\pm}$ just as we did for the tensor master variable. Recall that such $k_{\pm}$ may be just numbers or could contain a logarithm in the degenerate cases. The notions of vector Love number and magnetic susceptibility, though, are defined relative to the original fields, that is, the metric perturbation and Maxwell's vector potential \cite{Cardoso:2017cfl}. More precisely, vector Love numbers (magnetic susceptibility) measure the response of the BH when there is no electromagnetic (gravitational) tidal field at infinity. Physically, this can be thought of as the BH being perturbed by the presence of a massive yet neutral (light yet highly charged) companion. 

The decoupled degrees of freedom \eqref{dof} are defined up to their respective independent amplitudes, $\sigma_{(\pm)}$. These modulate the intensity with which each mode contributes to the total perturbation. Vanishing tidal fields at infinity are achieved for particular choices of such amplitudes. To see this, it is more convenient to trade the absolute amplitudes $\sigma_{(\pm)}$ by a relative amplitude $\boldsymbol{\Theta}$ and a global amplitude $\bold{A}$ defined as 
\beq
\boldsymbol{\Theta}&:=&\frac{\sigma_{(-)}}{\sigma_{(+)}}\,,\\
\bold{A}&:=&\frac{\left(\tilde{\Delta} +n^2-1\right)^{2}}{2 m_{V} (n-1) n \sin ^2(\epsilon )+\left(\tilde{\Delta} +n^2-1\right)^2}\frac{1}{\sigma_{(-)}}\,.
\eeq
In terms of these, the original fields\footnote{The condition on the gravitational perturbation is actually imposed on the gauge invariant variable $F_{a}$ in \eqref{gaugeinv}. For clarity here we give it in terms of the metric variable $h_{ai}$, but this is implicitly evaluated in the gauge $h_{T}^{(1)}=0$, where metric perturbation and gauge invariant variable coincide.} take the form 
\begin{widetext}
\begin{align}\label{met}
h_{ai}&=\bold{A} \frac{\epsilon_{ab}}{r^{n-2}}D^{b}\left[r^{n/2}z(r)^{\frac{2 l (n-1)+n-2}{2 (n-1)}}\left(k_{\rm vector}(\boldsymbol{\Theta})+\left(1+\frac{2(n-1)n\sin\epsilon}{\tilde{\Delta}+n^2-1}\boldsymbol{\Theta} \right)z^{-2l-1}+...\right)\right]\mathbb{V}_{i}\\\label{maxvec}
\delta A_{i}&=\bold{A}\boldsymbol{\Theta} \frac{\sqrt{n(n-1)}}{\kappa}r^{-\frac{n-2}{2}}z(r)^{\frac{2 l (n-1)+n-2}{2 (n-1)}}\left(k_{\rm magnetic}(\boldsymbol{\Theta})+\left(1-\frac{m_{V}\sin\epsilon}{\tilde{\Delta}+n^2-1}\frac{1}{\boldsymbol{\Theta}}\right)z^{-2l-1}+...\right)\mathbb{V}_{i}
\end{align}
\end{widetext}
where
\beq
k_{\rm vector}(\boldsymbol{\Theta})&=&k_{-}+\frac{2(n-1)n\sin\epsilon}{\tilde{\Delta}+n^2-1}\boldsymbol{\Theta} k_{+}\,,\\
k_{\rm magnetic}(\boldsymbol{\Theta})&=&k_{+}-\frac{m_{V}\sin\epsilon}{\tilde{\Delta}+n^2-1}\frac{k_{-}}{\boldsymbol{\Theta}}\,,\label{rsp}
\eeq
and we are keeping only the relevant terms of the master variables $\Psi_{\pm}$, that is, the tidal mode and the response fall-off. The quantities $k_{\rm vector}(\boldsymbol{\Theta})$ and $k_{\rm magnetic}(\boldsymbol{\Theta})$ are a measure of the response of the BH to a gravito-magnetic tidal field characterised by $\boldsymbol{\Theta}$, the relative intensity between the gravitational and magnetic contributions. The vector Love numbers and the magnetic susceptibility are precisely these quantities evaluated at the $\boldsymbol{\Theta}$'s in which there is no magnetic or no gravitational tidal fields respectively, that is, when no term $\sim z^{-2l-1}$ is present in the expansion of \eqref{maxvec} or \eqref{met} \cite{Cardoso:2017cfl},
\begin{align}
k_{\rm vector}=k_{-}+2\sin^{2}\epsilon\frac{m_{V}(n-1)n}{\left(\tilde{\Delta}+n^{2}-1\right)^{2}} k_{+}\label{lovevec}\,,\\
k_{\rm magnetic}=k_{+}+2\sin^{2}\epsilon\frac{m_{V}(n-1)n}{\left(\tilde{\Delta}+n^{2}-1\right)^{2}} k_{-}\label{magsuc}\,.
\end{align} 
With this, vector Love numbers $k_{\rm vector}$ and magnetic susceptibility $k_{\rm magnetic}$ are related simply by $+\leftrightarrow-$, and $k_{\pm}$ can be obtained from the master equations of $\Psi_{\pm}$ proceeding as we did for the tensor variable.

\subsubsection{Neutral and extremal limits}
It is convenient to deal first with neutral and extremal limits since the equations undergo a significant simplification. The response parameters for $Q=0$ were found in \cite{Hui:2020xxx} by solving \eqref{neutvect} with parameters \eqref{extneut1}-\eqref{extneut}\footnote{We obtain exact agreement with the results of \cite{Hui:2020xxx} with the exception of the magnetic susceptibility in the case that $l$ is a generic number. This may well be a typo and we take the opportunity to provide the corrected result:
\beq
k_{V}&=&(2\hat{L}+1)\frac{\Gamma\left(\hat{L}+1+\frac{1}{D-3}\right)^{2}\Gamma\left(1+\hat{L}-\frac{1}{D-3}\right)^{2}}{\Gamma\left(2\hat{L}+2\right)^{2}}\nonumber\\
&\times &\frac{\sin\left[\pi\left(\hat{L}+\frac{1}{D-3}\right)\right]\sin\left[\pi\left(\hat{L}-\frac{1}{D-3}\right)\right]}{\pi\sin\left(2\pi \hat{L}\right)}\,,
\eeq
where $\hat{L}$ and $k_{V}$ stand for our $l$ and $k_{\rm magnetic}$, respectively, in their notation.}. Let us consider a maximally charged BH $Q=M$. Static perturbations are described by the master variable $\psi_{\pm}$ (see \eqref{extmastvar}) subject to an hypergeometric equation with coefficients \eqref{extparam}. The response parameters $k_{\pm}$ in degenerate and non-degenerate cases are obtained as follows

\noindent {\bf First case:} $2l+1\notin \mathbb{N}$: The general solution can be written as \cite{Bateman:100233,Kristensson,NIST:DLMF}
\beq
\psi_{\pm}(z)&=& A F\left[a_{\pm},b_{\pm};c\vert z\right]\nonumber\\
&+&Bz^{1-c}F\left[a_{\pm}-c+1,b_{\pm}-c+1;2-c\vert z\right]\,,\label{genso}
\eeq
where $A$ and $B$ are arbitrary constants, $a_{\pm},b_{\pm}$ and $c$ are given in \eqref{extparam} and $F[a,b;c\vert z]$ denotes the hypergeometric function. Since the latter are normalised according to $F[a,b;c\vert 0]=1$, the response parameter $k_{\pm}$ enters the solution as (see Section \ref{DefLove})
%
\beq
\psi_{\pm}(z)&=&B k_{\pm} F\left[a_{\pm},b_{\pm};c\vert z\right]\,,\nonumber\\
&+&Bz^{1-c}F\left[a_{\pm}-c+1,b_{\pm}-c+1;2-c\vert z\right]\,.\label{genso1}
\eeq
%
Using the connection formula \cite{Bateman:100233,NIST:DLMF}
\begin{widetext}
\be
\frac{\sin\left[\pi\left(c-a-b\right)\right]}{\pi \Gamma\left(c\right)}F\left[a,b;c\vert z\right]=\frac{F\left[a,b;a+b-c+1\vert 1-z\right]}{\Gamma\left(c-a\right)\Gamma\left(c-b\right)\Gamma\left(a+b-c+1\right)}
-(1-z)^{c-a-b}\frac{F\left[c-a,c-b;c-a-b+1\vert 1-z\right]}{\Gamma\left(a\right)\Gamma\left(b\right)\Gamma\left(c-a-b+1\right)}\,,
\ee
\end{widetext}
one can write explicitly the analytic continuation of each hypergeometric function in \eqref{genso1} to a neighbourhood of $z=1$. In our case,
\begin{widetext}
\begin{align}\label{step}
\frac{\psi_{\pm}(z)}{B}=&-k_{\pm}\frac{\pi\Gamma(c)\left(1-z\right)^{c-a_{\pm}-b_{\pm}}F\left[c-a_{\pm},c-b_{\pm};c-a_{\pm}-b_{\pm}+1\vert 1-z\right]}{\sin\left[\pi\left(c-a_{\pm}-b_{\pm}\right)\right]\Gamma(a_{\pm})\Gamma(b_{\pm})\Gamma(c-a_{\pm}-b_{\pm}+1)}\\ \notag
&-\frac{\pi\Gamma(2-c)(1-z)^{c-a_{\pm}-b_{\pm}}z^{1-c}F\left[1-a_{\pm},1-b_{\pm};c-a_{\pm}-b_{\pm}+1\vert 1-z\right]}{\sin\left[\pi\left(c-a_{\pm}-b_{\pm}\right)\right]\Gamma(a_{\pm}-c+1)\Gamma(b_{\pm}-c+1)\Gamma(c-a_{\pm}-b_{\pm}+1)}\\\notag
&+(\text{Terms Regular at $z=1$})\,,
\end{align}
\end{widetext}
and using the further index displacement
\beq
&&z^{1-c}F\left[1-a_{\pm},1-b_{\pm};c-a_{\pm}-b_{\pm}+1\vert 1-z\right]\nonumber\\
&=&F\left[c-a_{\pm},c-b_{\pm};c-a_{\pm}-b_{\pm}+1\vert 1-z\right]\,,
\eeq
equation \eqref{step} reads
\beq
&&\frac{\psi_{\pm}(z)}{B}=-\left[\frac{k_{\pm}\Gamma(c)}{\Gamma(a_{\pm})\Gamma(b_{\pm})}+\frac{\Gamma(2-c)}{\Gamma(a_{\pm}-c+1)\Gamma(b_{\pm}-c+1)}\right]\nonumber\\
&\times& \pi \frac{\left(1-z\right)^{c-a_{\pm}-b_{\pm}}F\left[c-a_{\pm},c-b_{\pm};c-a_{\pm}-b_{\pm}+1\vert 1-z\right]}{\sin\left[\pi\left(c-a_{\pm}-b_{\pm}\right)\right]\Gamma(c-a_{\pm}-b_{\pm}+1)}\nonumber\\
&+&(\text{Terms Regular at $z=1$})\,.\label{step2}
\eeq
The coefficients \eqref{extparam} of the extremal master equation satisfy
\begin{equation}
c-a_{\pm}-b_{\pm}=-\frac{\Sigma_{\pm}}{n-1}\,,
\end{equation}
so the first term in \eqref{step2} is singular at the horizon $z=1$ unless $k_{\pm}$ are 
chosen to make the prefactor vanish, that is, in terms of $l$ and the symbol $S_{(\pm,\pm)}$ (see \eqref{symbol}), 
\beq
k_{\pm}&=&-\frac{\left(S_{(\pm,+)}+l\right)\left(S_{(\pm,-)}+l\right)}{2l(2l+1)}\frac{\Gamma(-2l)}{\Gamma(2l)}\nonumber\\
&\times&\frac{\Gamma\left(S_{(\pm,+)}+l\right)}{\Gamma\left(S_{(\pm,+)}-l\right)}\frac{\Gamma\left(S_{(\pm,-)}+l\right)}{\Gamma\left(S_{(\pm,-)}-l\right)}\,.
\eeq

\noindent {\bf Second case:} $2l+1\in\mathbb{N}$: Here we shall additionally distinguish between $D\neq4$ and $D=4$. In the former case the general solution takes the form 
\beq
\psi_{\pm}(z)&=&A F\left[a_{\pm},b_{\pm}; c\vert z\right]\nonumber\\
&+&B F\left[a_{\pm},b_{\pm}; a_{\pm}+b_{\pm}-c+1\vert 1-z\right]\,,
\eeq
and only the second solution is regular at $z=1$, which implies $A=0$. Again using appropriate connection formulas in the degenerate cases it is easy to show that \cite{Bateman:100233,Kristensson,NIST:DLMF}
\beq
&&F\left[a_{\pm},b_{\pm}; a_{\pm}+b_{\pm}-c+1\vert 1-z\right]\nonumber\\
&\sim&\left(z^{-2l-1}+...+ R_{\pm} F\left[a_{\pm},b_{\pm};c\vert z\right] \ln z\right)\,,
\eeq
where the ellipsis denotes subleading terms in $z$ and, in terms of $l$ and $S_{(\pm,\pm)}$, $R_{\pm}$ reads  
\beq
R_{\pm}&=&(-1)^{2l}\frac{\left(S_{(\pm,+)}+l\right)\left(S_{(\pm,-)}+l\right)}{(2l+1)!(2l)!}\frac{\Gamma\left(S_{(\pm,+)}+l\right)}{\Gamma\left(S_{(\pm,+)}-l\right)}\nonumber\\
&\times&\frac{\Gamma\left(S_{(\pm,-)}+l\right)}{\Gamma\left(S_{(\pm,-)}-l\right)}\,.
\eeq
If $D=4$, however, the coefficients in \eqref{extparam} become highly degenerate,
\begin{equation}
a_{+}=a_{-}+2=c+3,\ \ \ b_{+}=b_{-}+2=c-4,\ \ \ c=2(l+1)\,.
\end{equation}
In particular, all of them are integers and the general solution is
\begin{equation}
\psi_{\pm}(z)=A F\left[a_{\pm},b_{\pm};c\vert z\right]+Bz^{-2l-1}\begin{cases}F\left[4,-1;-2l\vert z\right] & (+) \\ F\left[2,-3;-2l\vert z\right] & (-)\end{cases}\,.
\end{equation}
Regularity at the horizon $z=1$ sets $A=0$ and the functions in the braces are just polynomials in $z$ (we recall that $L=1$ has no gravitational mode $(-)$). This is a purely tidal field and, thus, we conclude that in $D=4$
\begin{equation}
k_{\pm}=0\,.
\end{equation}
To summarise, we have found that the response parameters $k_{\pm}$ of the extremal BHs are given by
\begin{widetext}
\begin{equation}\label{resultextremal}
k_{\pm}=\begin{cases}-\frac{\left(S_{(\pm,+)}+l\right)\left(S_{(\pm,-)}+l\right)}{2l(2l+1)}\frac{\Gamma{\left(-2l\right)}}{\Gamma{\left(2l\right)}}\frac{\Gamma\left(S_{(\pm,+)}+l\right)}{\Gamma\left(S_{(\pm,+)}-l\right)}\frac{\Gamma\left(S_{(\pm,-)}+l\right)}{\Gamma\left(S_{(\pm,-)}-l\right)}&2l+1\notin\mathbb{Z}\\ (-1)^{2l}\frac{\left(S_{(\pm,+)}+l\right)\left(S_{(\pm,-)}+l\right)}{(2l+1)!2l!}\frac{\Gamma\left(S_{(\pm,+)}+l\right)}{\Gamma\left(S_{(\pm,+)}-l\right)}\frac{\Gamma\left(S_{(\pm,-)}+l\right)}{\Gamma\left(S_{(\pm,-)}-l\right)} \ln z &2l+1\in\mathbb{Z}, \ \ D\ne4 \\ 0 & D=4\end{cases}\,
\end{equation}
\end{widetext}
where the symbol $S_{(\pm,\pm)}$ is defined in \eqref{symbol}. The vector Love numbers and the magnetic susceptibility are obtained by plugging such $k_{\pm}$'s into \eqref{lovevec} and \eqref{magsuc}, respectively.

It is worth making a remark here before considering the BH with finite $Q$ and $T_{H}$. We have found that vector Love numbers and magnetic susceptibilities do not vanish at extremality unless $D=4$. This is in contrast with the tensor sector, where Love numbers are $\sim T_{H}^{2l+1}$ and thus vanish at zero temperature. Quite the opposite, for vector perturbations the charge triggers polarisations in modes that are otherwise not excited in the neutral case. Indeed, Ref.~\cite{Hui:2020xxx} found that some special modes in the vector sectors (both of gravitational and electromagnetic types) do not exhibit a static response to external fields when the BH is not charged. These have $l\in\frac{1}{2}\mathbb{N}$ in $D=5$ or $L=N(D-3)\pm1$ in $D>5$ with $N\in\mathbb{N}$ (notice these always include the special mode $L=1$). Such special modes seem to be a property of magnetic-like perturbations since they have no analogue in the corresponding scalar sectors. However, when the BH is maximally charged we have found that such harmonics do not fall within any special class and, therefore, exhibit some polarisation (of both gravitational and magnetic types) according to \eqref{resultextremal}. Therefore, a non-trivial static response in these harmonics is a signature of non-vanishing charge. In the following section we show that, indeed, charging up the BH has the effect (in the vector sector) of increasing the intensity of the response and even turning on new modes of polarisation.

\subsubsection{Finite charge and temperature}
For intermediate values of the BH charge the equation governing static perturbations in the vector sector \eqref{mastertensor} has an extra pole due to the Cauchy horizon. Thus, the treatment in terms of hypergeometric functions considered in the neutral and extremal cases does not apply. While in the degenerate case $2l+1\in \mathbb{N}$ it is still possible to obtain exact analytic results, for general $l$ with $2l+1\notin \mathbb{N}$ we proceed numerically.

\begin{figure}[t!]
		\includegraphics[width=0.45\textwidth]{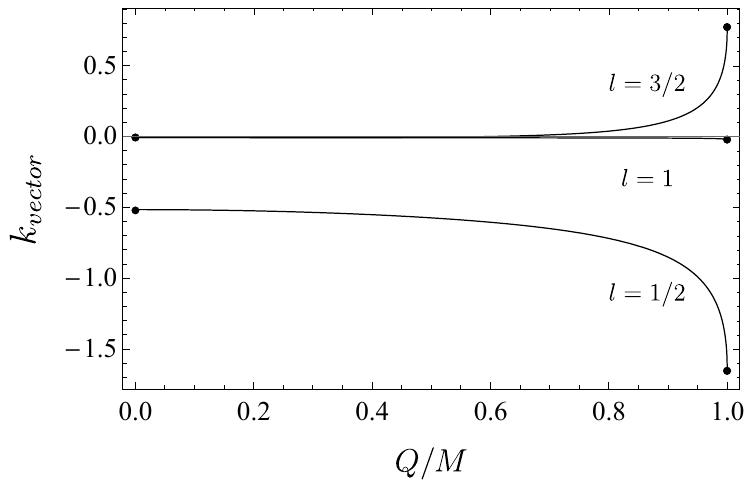}
		\includegraphics[width=0.45\textwidth]{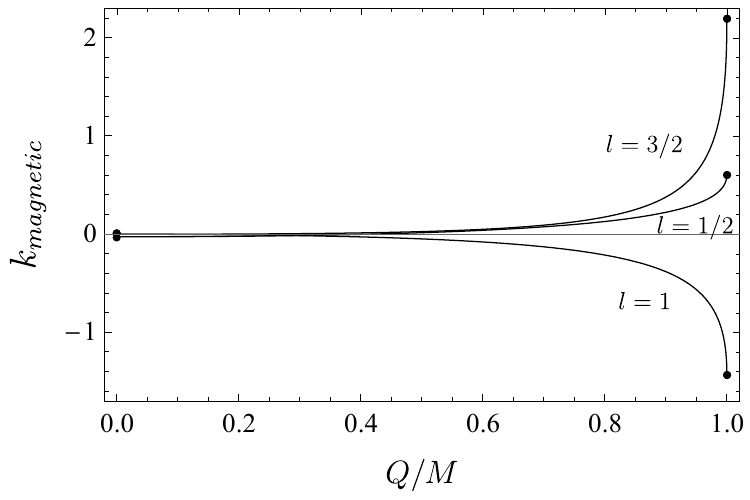}
		\caption{$k_{\rm vector}$ (top) and $k_{\rm magnetic}$ (bottom), in the degenerate case $2l+1\in\mathbb{N}$ (omitting the factor $\ln z$). We show $L=4,8,12$ in $D=11$. Solid black lines are the analytic results obtained as explained in the main text. These interpolate exactly between the analytic predictions at $Q=0$ and $Q=M$, represented with solid black dots.} \label{fig:Running}
\end{figure}

\noindent {\bf First case:} $2l+1\in \mathbb{N}$: Again we shall distinguish the cases $D\ne4$ and $D=4$. Consider first $D\ne4$ and let $L_{\alpha,\beta,\gamma,\delta,\eta,h,z_{c}}[\cdot]$ be Heun's operator, so that Heun's equation for a function $f(z)$ reads $L_{\alpha,\beta,\gamma,\delta,\eta,h,z_{c}}[f(z)]=0$. Much like in the tensor case, after choosing the normalisation of $\Psi_{\rm tidal(\pm)}(z)$ as $\Psi_{\rm tidal(\pm)}(z)=1+O(z)$, imposing Heun's equation \eqref{mastertensor} on the second solution of \eqref{gensol} fixes $R_{\pm}$ completely. In particular, using that $L_{\alpha,\beta,\gamma,\delta,\eta,h,z_{c}}[\Psi_{\rm resp(\pm)}(z)]=0$ and expanding $L_{\alpha,\beta,\gamma,\delta,\eta,h,z_{c}}[z^{-2l-1}\Psi_{\rm tidal(\pm)}(z)]=z^{-2l-2}\sum_{i=0}a^{(\pm)}_{i}z^{i}$ it follows that $R_{\pm}$ is formally given by
\begin{equation}\label{Rs}
R_{\pm}=-\frac{a^{(\pm)}_{2l}}{2l+1}\,.
\end{equation}
The coefficient $a^{(\pm)}_{2l}$ depends on the coefficients at all previous orders $a^{(\pm)}_{i<2l}$, and it is not clear whether it is possible to give the general result for any $l,n,$ and $\epsilon$ (as it was in the tensor sector). However, given a particular value of $l$ one can just solve all previous orders $a_{i<2l}$ and get, through \eqref{Rs}, the exact result of $R_{\pm}$ in terms of $n,\epsilon$. For example, for $l=1$ we find
\begin{widetext}
\beq
R_{\pm}&=&\left[-2 n^4+3 n^3-7 n^2+11 n-13\pm\left(-2 n^2+3 n+11\right) \tilde{\Delta} +\left(2 n^2-3 n+1\right) \left(n^2\pm\tilde{\Delta} -7\right) \cos (2 \epsilon )\right]\nonumber\\ 
&\times&\frac{n^2 \sec ^6\left(\frac{\epsilon }{2}\right)}{96 (n-1)^6}\,,
\eeq
\end{widetext}
and it is easy to check that this interpolates between the neutral result in \cite{Hui:2020xxx} and the extremal one in \eqref{resultextremal} (as $\epsilon$ goes from $0$ to $\pi/2$, respectively). Love numbers and magnetic susceptibilities are finally obtained by plugging these results into \eqref{lovevec} and \eqref{magsuc}. In Figure~\ref{fig:Running} we show $k_{\rm vector}$ and $k_{\rm magnetic}$ in $D=11$ for several harmonics $l$.
Next we consider $D=4$. Once again we find a second solution with $R_{\pm}=0$ and $\Psi_{\rm tidal(\pm)}(z)$ a polynomial of degree $<2l+1$. This is the solution that is regular at the horizon $z=1$ and consists solely of a tidal field, so once more $k_{\rm vector}=0$ and $k_{\rm magnetic}=0$ in four dimensions, now for any value of the BH temperature $T_{H}$.

\begin{figure}[t!]
\includegraphics[width=0.45\textwidth]{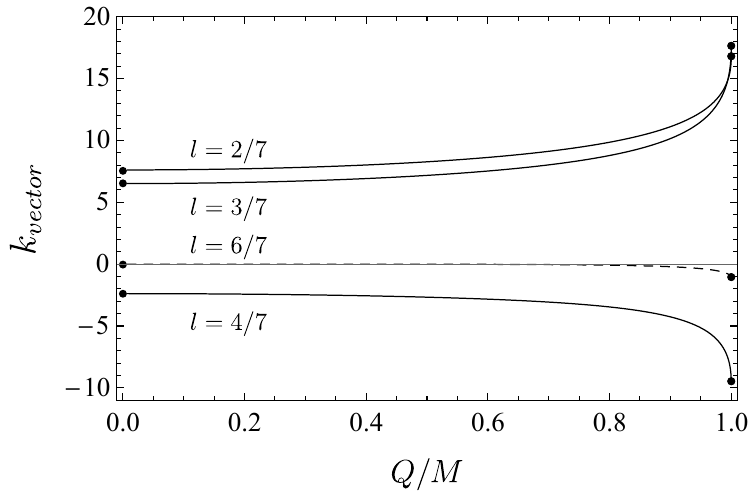}
\includegraphics[width=0.45\textwidth]{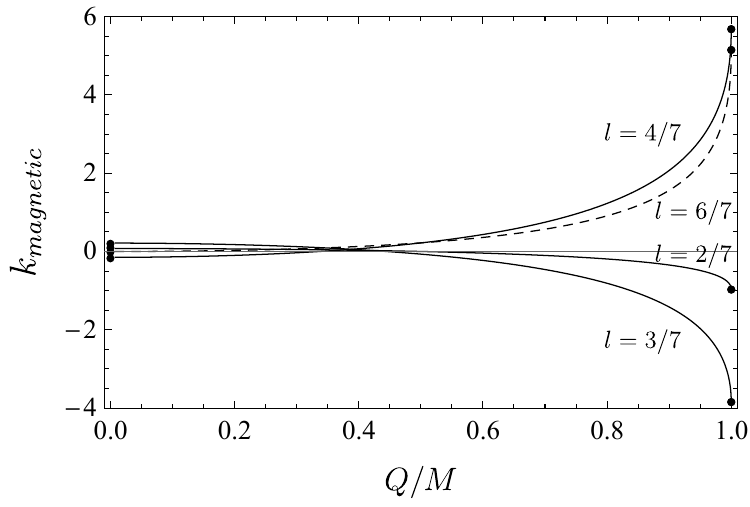}
\caption{$k_{\rm vector}$ (top) and $k_{\rm magnetic}$ (bottom), in the general case $2l+1\notin\mathbb{N}$. We show $L=2,3,4,6$ in $D=10$. Solid black lines are the results obtained numerically and solid black dots are the analytical results at $Q=0$ and $Q=M$. The harmonic $L=6$, represented with a dashed line, is an example of the special modes that do not polarise at $Q=0$, but exhibit a non-trivial response as $Q$ grows.} \label{fig:General}
\end{figure}

\noindent {\bf Second case:} $2l+1\notin \mathbb{N}$: In this case, there seems to be no clear way of guessing the results for $k_{\pm}$ out of those for $R_{\pm}$ in \eqref{Rs} as we did for the tensor sector. Thus, we proceed numerically by implementing a standard shooting method (similar to that used in \cite{Cano:2021qzp}) which matches the regular solution at the horizon $z=1$ with one at infinity of the form \eqref{enn}, thus obtaining the values of $k_{\pm}$. Then $k_{\rm vector}$ and $k_{\rm magnetic}$ follow from \eqref{lovevec} and \eqref{magsuc}. In Figure \ref{fig:General} we show $k_{\rm vector}$ and $k_{\rm magnetic}$ in $D=10$ for several harmonics $l$.

These results confirm the analytical predictions at $Q=0$ and $T_{H}=0$. We can conclude that charged BHs exhibit a stronger response to gravitational and electromagnetic tidal fields, relative to their neutral counterparts. Even more, a non-vanishing charge turns on new modes of gravitational and magnetic polarisation that are otherwise not responsive for $Q=0$. A non-trivial static response in such harmonics is, therefore, a definite signature of charge. We can also confirm that in four dimensions, for all $T_{H}$, both tidal Love numbers and magnetic susceptibilities vanish. This property is strongly related to the fact that, in $D=4$, the equations become degenerate enough so as to admit purely-growing polynomial solutions.
%

\section{Discussion}

We have studied the effect of charge on the static polarizability of BHs in $D\geq4$ spacetime dimensions. While the four-dimensional setup remains intriguingly special, with all response parameters vanishing, TLNs and magnetic susceptibilities exhibit a rich structure in $D>4$. In particular, charging up the BH turns on new vector-type modes of polarisation, while tensor Love numbers (encoding also the response to scalar tidal fields) decrease and eventually vanish at extremality. More precisely, our results can be summarised as follows.\\
\noindent \textbf{(i)} The relevant differential equations are of Fuchsian type with 4 poles (Heun) at infinity, the event and Cauchy horizons and the curvature singularity. In the neutral ($Q=0$) and extremal ($T_{H}=0$) limits, the equations become hypergeometric and TLNs are exactly solvable.\\ 
\noindent \textbf{(ii)} For the tensor sector of gravitational perturbations we showed that all TLNs (equivalently, the response to scalar tidal fields) vanish at extremality, $T_{H}=0$, and confirmed the results in the literature for $Q=0$~\cite{Hui:2020xxx}. Even for arbitrary (subextremal) values of the BH charge we are able to obtain the exact result analytically, finding that tensor TLNs follow a power law in the BH temperature, $k_{\text{tensor}}\sim T_{H}^{2l+1}$.\\
\noindent \textbf{(iii)} For the so-called vector sector, we find analytical expressions for the Love numbers and magnetic susceptibilities at extremality, $T_{H}=0$. We also recover results in the literature at zero charge~\cite{Hui:2020xxx}, correcting the reported result for the magnetic susceptibilities. For intermediate $Q$'s we find some results analytically and some numerically, in all cases confirming our analytic predictions at $T_{H}=0$ and those in the literature for the neutral case, $Q=0$. In contrast to the tensor sector, we found that charged BHs exhibit a stronger response to gravitational and electromagnetic tidal fields (of vector type), relative to their neutral counterparts. In addition, we showed that the BH charge excites new modes of gravitational and magnetic polarisation that are otherwise not responsive for $Q=0$. A non-trivial static response in such harmonics is, therefore, a definite signature of charge. \\
\noindent \textbf{(iv)} Our results show that in four dimensions and for all values of the charge, all response parameters vanish. This property is strongly related to the fact that, in $D=4$, the equations become degenerate enough so as to admit purely-growing polynomial solutions.

Our results raise interesting questions in various directions. First, it is desirable to understand and explore further the special properties of tensor modes (scalar fields) at extremality, possibly including black hole rotation in a suitable spin configuration. In parallel, it would also be interesting to consider 
BHs carrying a more general charge configuration and study whether these excite new modes of polarisation, similarly to what we found in the vector sector. These and more aspects about tidal deformability of charged BHs will be addressed in future work.

\section*{Acknowledgements}
D. P. gratefully acknowledges the hospitality of the group at CENTRA. D. P. thanks Francisco Duque, Rodrigo Vicente, Laura Bernard, Marc Casals and Alexandre Le Tiec for interesting conversations. 
D. P. is funded in part by a Centro de Excelencia Internacional UAM/CSIC FPI pre-doctoral grant. 
V. C. is a Villum Investigator supported by VILLUM FONDEN (grant no. 37766) and a DNRF Chair supported by the Danish National Research Foundation.
This project has received funding from the European Union's Horizon 2020 research and innovation programme under the Marie Sklodowska-Curie grant agreement No 101007855.
We thank FCT for financial support through Project~No.~UIDB/00099/2020.
We acknowledge financial support provided by FCT/Portugal through grants PTDC/MAT-APL/30043/2017 and PTDC/FIS-AST/7002/2020.
The authors would like to acknowledge networking support by the GWverse COST Action CA16104, ``Black holes, gravitational waves and fundamental physics''.


\bibliography{Gravities} 

\end{document}